\documentclass[12pt]{article}
\begin{document}
\title{Strings and Planck Oscillators}
\author{B.G. Sidharth\\
International Institute for Applicable Mathematics \& Information Sciences\\
Hyderabad (India) \& Udine (Italy)\\
B.M. Birla Science Centre, Adarsh Nagar, Hyderabad - 500 063 (India)}
\date{}
\maketitle
\begin{abstract}
We compare and contrast two theories which both start in one dimension at the Planck scale, viz., Quantum SuperString or M-Theory and a theory of Planck oscillations.
\end{abstract}
\section{Introduction}
String Theory has evolved into M-Theory over the decades and is receiving a considerable amount of attention these days. Less well known is a theory based on oscillators at the Planck scale which has lead to a number of very interesting results. In both cases we start with one dimensional phenomena at the Planck scale, vibrations in the case of strings and oscillators in the latter case. But from this point on the theories take different routes. In what follows we will briefly study, compare and contrast these approaches.
\section{String Theory}
Inspite of great success, the standard theory has failed to quantize gravitation. One of the obstacles has been the point spacetime concept ingrained in these theories. For the past few decades Quantum Gravity schemes as also string theory have attempted to break out of this limitation. Let us first consider string theory.\\
We begin with the important work of T. Regge in the fifties \cite{reg,roma,tassie}, in which he mathematically analysed using techniques like analytically continuing the angular momentum into the complex plane, particle resonances. These resonances seem to fall along a straight line plot, with the angular momentum being proportional to the square of the mass.
\begin{equation}
J \propto M^2,\label{ex}
\end{equation}
All this suggested that resonances had angular momentum, on the one hand and resembled extended objects, that is particles smeared out in space. For example, mathematically this was like two heavy objects attached to the two ends of a string, or a rotating stick.\\
This went contrary to the belief that truly elementary particles were points in space. Infact at the turn of the twentieth century, Poincare, Lorentz, Abraham and others had toyed with the idea that the electron had a finite extension, but they had to abandon this approach, because of a conflict with Special Relativity. The problem is that if there is a finite extension for the electron then forces on different parts of the electron would exhibit a time lag, requiring the so called Poincare stresses for stability \cite{rohr,barut,feynman}.\\
In this context, it may be mentioned that in the early 1960s, Dirac came up with an imaginative picture of the electron, not so much as a point particle, but rather a tiny closed membrane or bubble. Further, the higher energy level oscillations of this membrane would represent the ``heavier electrons'' like muons \cite{dirac}.\\
Then, in 1968, G. Veneziano came up with a unified description of the Regge resonances (\ref{ex}) and other scattering processes. Veneziano considered the collision and scattering process as a black box and pointed out that there were in essence, two scattering channels, $s$ and $t$ channels. These, he argued gave a dual description of the same process \cite{ven,venezia}.\\
In an $s$ channel, particles A and B collide, form a resonance which quickly disintegrates into particles C and D. On the other hand we have in a $t$ channel scattering  particles A and B approach each other, and interact via the exchange of a particle $q$. The result of the interaction is that particles C and D emerge. If we now enclose the resonance and the exchange particle $q$ in an imaginary black box, it will be seen that the $s$ and $t$ channels describe the same input and the same output: They are essentially the same.\\
There is another interesting hint which we get from Quantum Chromo Dynamics. Let us come to the inter-quark potential \cite{lee,smolin}. There are two interesting features of this potential. The first is that of confinement, which is given by a potential term like
$$V (r) \approx \sigma r, \quad r \to \infty ,$$
where $\sigma$ is a constant. This describes the large distance behavior between two quarks. The confining potential ensures that quarks do not break out of their bound state, which means that effectively free quarks cannot be observed.\\
The second interesting feature is asymptotic freedom. This is realized by a Coulumbic potential
$$V_c (r) \approx - \frac{\propto (r)}{r} (\mbox{small}\, r)$$
$$\mbox{where}\, \propto (r) \sim \frac{1}{ln(1/\lambda^2r^2)}$$
The constant $\sigma$ is called the string tension, because there are string models which yield $V(r)$. This is because, at large distances the inter-quark field is string like with the energy content per unit length becoming constant. Use of the angular momentum - mass relation indicates that $\sigma \sim (400 MeV)^2$.\\
Such considerations lead to strings which are governed by the equation \cite{walter,fog,bgskluwer,st}
\begin{equation}
\rho \ddot {y} - T y'' = 0,\label{e1}
\end{equation}
\begin{equation}
\omega = \frac{\pi}{2l} \sqrt{\frac{T}{\rho}},\label{e2}
\end{equation}
\begin{equation}
T = \frac{mc^2}{l}; \quad \rho = \frac{m}{l},\label{e3}
\end{equation}
\begin{equation}
\sqrt{T/\rho} = c,\label{e4}
\end{equation}
$T$ being the tension of the string, $l$ its length and $\rho$ the line density and $\omega$ in (\ref{e2}) the frequency. The identification (\ref{e2}),(\ref{e3}) gives (\ref{e4}), where $c$ is the velocity of light, and (\ref{e1}) then goes over to the usual d'Alembertian or massless Klein-Gordon equation. (It is worth noting that as $l \to 0$ the potential energy which is $\sim \int^l_0 T (\partial y/\partial x)^2 dx$ rapidly oscillates.)\\
Further, if the above string is quantized canonically, we get
\begin{equation}
\langle \Delta x^2 \rangle \sim l^2.\label{e5}
\end{equation}
The string effectively shows up as an infinite collection of harmonic oscillators \cite{fog}. It must be mentioned that (\ref{e5}) and (\ref{e2}) to (\ref{e4}) both show that $l$ is of the order of the Compton wavelength. This has been called one of the miracles of string theory by Veneziano \cite{veneziano}. In fact the minimum length $l$ turns out to be given by $T/\hbar^2 = c/l^2$, which from (\ref{e3}) and (\ref{e4}) is seen to give the Compton wavelength.\\
This is a description of what may be called a ``Bosonic String''. These theories have certain technical problems, for example they allow the existence of tachyons. Further they do not easily meet the requirements of Quantum theory, as for example the commutation relations. The difficulties are resolved only in twenty six dimensions.\\
If the relativistic quantized string is given rotation \cite{ramond}, then we get back the equation for the Regge trajectories given in (\ref{ex}) above. Here we are dealing with objects of finite extension rotating with the velocity of light rather like spinning black holes. It must be pointed out that, in superstring theory, there is an additional term $a_0$
\begin{equation}
J \leq (2\pi T)^{-1} M^2 + a_0 \hbar, \, \mbox{with}\, a_0 = +1 (+2) \, \mbox{for \, the \, open\, (closed)\, string.}\label{e6}
\end{equation}
In equation (\ref{e6}) $a_0$ comes from a zero-point energy effect. When $a_0 = 1$ we have the usual gauge bosons and when $a_0 = 2$ we have the gravitons.\\
The theory of Quantum Super Strings in contrast requires only ten dimensions. Here, Quantum operators describing anti-commuting variables satisfy anti-commutation relations. Indeed this bivalence is a hallmark of supersymmetry itself.\\
The extra dimensions that appear in String theories reduce to the four dimensions of the physical spacetime by virtue of the fact that the redundant dimensions are treated as curled up into a negligible extension, in the manner suggested by Kaluza and later Klein in the early twentieth century. Kaluza's original motivation had been to unify electromagnetism and gravitation by introducing a fifth negligible coordinate. The curling up takes place at the Planck scale \cite{kaluza}.\\
A finite extension for an elementary particle, as in String theories can be shown to lead to new commutation relations, as was done by Snyder in the forties. In this case two space coordinates like $x$ and $y$ do not commute. Snyder's original motivation had been to fudge and eliminate singularities and divergences in Quantum fields. We remark that what this implies is that space coordinates in some sense take on the character of momenta in addition, though this happens at very small scales or high energies. Effectively there is a modification of the Uncertainty Principle
\begin{equation}
\Delta x \geq \frac{\hbar}{\Delta_P} + l^2 \frac{\Delta_P}{\hbar}\label{e7}
\end{equation}
What all this means is we cannot go down to lower and lower scales arbitrarily. As we approach the minimum length we return to the larger universe \cite{witten}. We will return to this point a little later.\\
The interesting thing about Quantum Superstring theory is the natural emergence of the spin 2 graviton as can be seen from (\ref{e6}), or as Witten puts it, the theory ``predicts'' gravitation.\\
Meanwhile Supersymmetry or SUSY developed in parallel. This theory requires that each particle with integral spin has a counterpart with the same mass but having half integral spin. That is Bosons have their supersymmetric counterparts in Fermions. SUSY is then broken so that the counterparts would have a much greater mass, which would then account for the fact that these latter have not been observed. Nevertheless the fact that in this theory gravitation can be unified with the other forces makes it attractive.\\
Infact this had lead to Supergravity in which the spin 2 graviton has the spin 3/2 counterpart, the gravitino. Supergravity requires eleven spacetime dimensions, one more than Superstring theory.\\
Unfortunately Supergravity began to fade from the mid eighties because of the fact that, as shown by Witten and others, handedness cannot easily emerge on reduction to the four physical spacetime dimensions from eleven. On the other hand the Quantum Super String theory was in comparison altogether more satisfactory. We could say that the earlier Bosonic String theory worked in a spacetime that was Bosonic, there being no place for spin. QSS works in a Fermionic spacetime where we have the modification (\ref{e7}).\\
So in the mid eighties ten dimensional QSS displaced Supergravity. There were five QSS theories - $E_8 \times E_8$ heterotic, $SO(32)$ heterotic, the Type I, the Type IIA and Type IIB. Of these the Type I is an open string while the others form closed loops. The $E_8 \times E_8$ appeared to explain many features of elementary particles and their forces.\\
However there were some disturbing questions. Why were there five different theories? After all we need a unique theory. And then why ten dimensions, while Supersymmetry allows eleven dimensions? Another not very convincing factor was the fact that particles were being represented as one dimensional strings. Surely a more general formulation as noted above would have two dimensional surfaces or membranes or even p-dimensional entities which we may call p-branes. This generalization resembles the earlier attempt of Dirac's representing particles as a shell or membrane. Infact if the radius of the circle shrinks, the mebrane begins to resemble a rolled up object in ten dimensions. It reduces to a Type IIA Superstring.\\
In such deformations certain topological properties can remain conserved. A good example is a knot in a set of field lines. Such knots or solitons remain as such and exhibit a particle type behaviour. A magnetic monopole can be characterized in this way, that is as a twisted knot of magnetic lines. It can be said to carry a topological charge. This is to be contrasted with the charges carried by particles like electrons and quarks which can be put within the framework of the Noether Conservation Theorem. In this context an interesting conjecture is that of Montonen and Olive \cite{olive}: There could be a dual formulation in which the roles of the usual charges and topological charges are reversed. In such a formulation for example a particle with charge $e$ would show up as a soliton with charge $\frac{1}{e}$.\\
Over the past few years, a variant called $M$ Theory arising from these generalizations has attracted much attention. This theory also uses Supersymmetry, which is broken so that the postulated particles do not have the same mass as the known particles. Further these new masses must be much too heavy to be detected by current accelerators. The advantage of Supersymmetry is that a framework is now available for the unification of all the interactions including gravitation. It may be mentioned that under a SUSY transformation, the laws of physics are the same for all observers, which is the case in General Relativity (Gravitation) also. Under SUSY there can be a maximum of eleven dimensions, the extra dimensions being curled up as in Kaluza-Klein theories. In this case there can only be an integral number of waves around the circle, giving rise to particles with quantized energy. However for observers in the other four dimensions, it would be quantized charges, not energies. The unit of charge would depend on the radius of the circle, the Planck radius yielding the value $e$. This is the root of the unification of electromagnetism and gravitation in these theories.\\
The relevance of all this is that p-branes can be characterized as solitons. For example a ten dimensional string can show up as a p-brane with $p = 5$. In this case a strongly interacting string would be the dual of a weakly interacting 5-brane. In 1990 the Montonen-Olive duality which was between electricity and magnetism in ordinary four dimensional space, was generalized to four dimensional Superstrings.\\
This duality was called S-duality, to distinguish it from the well known T-duality which relates two kinds of particles that arise when the string loops around by a compact dimension: There would be vibrations on the one hand and multiple windings on the other. Winding particles over a circle of radius $r$ correspond to vibrating particles in a circle of radius $1/r$ and conversely on the lines of (\ref{e7}). Such a behaviour is characteristic of minimum spacetime intervals. In this picture the solitonic interaction is given by the reciprocal of the string interaction, in confirmity with the Montonen-Olive conjecture.\\
A further interesting development was the realization that in the reduction of the dimensions of spacetime to four dimensions the string and the corresponding soliton each acquire a T-duality. Moreover the T-duality of the solitonic string is the S-duality of the fundamental string and conversely. We have here a duality of dualities. It also implies that the interaction charges in one universe show up as sizes in the dual.\\
Further the eleventh and extra dimension of the M-Theory could be shrunk, so that there would be two ten dimensional universes connected by the eleven dimensional spacetime. Now particles and strings would exist in the parallel universes which can interact through gravitation. The interesting aspect of the above scenario is that it is possible to concieve of all the four interactions converging at an energy far less than the Planck energy $(10^{19}GeV$). Infact the Planck energy is so high that it is beyond forseeable experiments. Thus this would bring the eleven dimensional M-Theory closer to experiment. There have been further developments involving what are called Dirichlet surfaces. It is now suspected that black holes can be treated as intersecting black branes wrapped around seven curled up dimensions. There is here, an interesting interface between M-Theory and black hole physics \cite{green}. In M-Theory, the position coordinates become matrices and this leads to, as we will see, a noncommutative geometry or fuzzy spacetime in which spacetime points are no longer well defined \cite{madore}
$$[x,y] \ne 0$$
From this point of view the mysterious $M$ in M-Theory could stand for Matrix, rather than Membrane.
 In any case, as we will argue, fuzzy spacetime may well hold the key for the unification of all interactions.\\
So M-Theory is the new avatar of QSS. Nevertheless it is still far from being the last word. There are still any number of routes for compressing ten dimensions to our four dimensions. There is still no contact with experiment. It also appears that these theories lead to an unacceptably high cosmological constant and so on.
\section{Planck Oscillators}
As we have seen, and this as noted being true in Quantum Gravity as well as in Quantum
Super String Theory, we encounter phenomena at a minimum scale. It is well known, and this was realized by Planck himself, that there is an absolute minimum scale in the universe, and this is,
$$l_P = \left(\frac{\hbar G}{c^3}\right)^{\frac{1}{2}} \sim
10^{-33}cm$$
\begin{equation}
t_P = \left(\frac{\hbar G}{c^5}\right)^{\frac{1}{2}} \sim
10^{-42}sec\label{ea1}
\end{equation}
Yet what we encounter in the real world is, not the Planck scale,
but the elementary particle Compton scale. The explanation given for
this is that the very high energy Planck scale is moderated by the
Uncertainty Principle. The question which arises is, exactly how
does this happen? We will now present an argument to show how the
Planck scale leads to the real world Compton scale, via
fluctuations and the modification of the Uncertainty Principle.\\
We note that (\ref{ea1}) defines the absolute minimum physical scale
\cite{bgsafdb,rr1,rr2,garay}. Associated with (\ref{ea1}) is the Planck mass
\begin{equation}
m_P \sim 10^{-5}gm\label{ea2}
\end{equation}
There are certain interesting properties associated with
(\ref{ea1}) and (\ref{ea2}). $l_P$ is the Schwarzschild radius of a
black hole of mass $m_P$ while $t_P$ is the evaporation time for
such a black hole via the Beckenstein radiation \cite{rr3}.
Interestingly $t_P$ is also the Compton time for the Planck mass,
a circumstance that is symptomatic of the fact that at this scale,
electromagnetism and gravitation become of the same order. Indeed all this fits in very well with Rosen's analysis
that such a Planck scale particle would be a mini universe
\cite{rr5,rr6}. We will now invoke a time varying gravitational
constant
\begin{equation}
G \approx \frac{lc^2}{m\sqrt{N}} \propto (\sqrt{N}t)^{-1} \propto
T^{-1}\label{ea3}
\end{equation}
which resembles the Dirac cosmology and features in another scheme to be discussed elsewhere (Cf.ref.\cite{walter}),
in which (\ref{ea3}) arises due to the fluctuation in the particle
number \cite{uof,rr8,rr9,rr10}. In (\ref{ea3}) $m$ and $l$ are the
mass and Compton wavelength of a typical elementary particle like
the pion while $N \sim 10^{80}$ is the number of elementary
particles in the universe, and
$T$ is the age of the universe.\\
In this scheme wherein (\ref{ea3}) follows from the theory, we use
the fact that given $N$ particles, the fluctuation in the particle
number is of the order $\sqrt{N}$, while a
typical time interval for the fluctuations is $\sim \hbar /mc^2$,
the Compton time. We will come back to this point later. So anticipating later work we have
$$\frac{dN}{dt} = \frac{\sqrt{N}}{\tau}$$
whence on integration we get,
$$T = \frac{\hbar}{mc^2} \sqrt{N}$$
and we can also deduce its spatial counterpart, $R = \sqrt{N} l$,
which is the well known empirical Eddington formula. We will return to this later.\\
Equation (\ref{ea3}) which is an order of magnitude relation is
consistent with observation \cite{rr11,melnikov} while it may be remarked
that the Dirac cosmology itself has inconsistencies.\\
Substitution of (\ref{ea3}) in (\ref{ea1}) yields
$$l = N^{\frac{1}{4}} l_P,$$
\begin{equation}
t = N^{\frac{1}{4}} t_P\label{ea4}
\end{equation}
where $t$ as noted is the typical Compton time of an elementary
particle. We can easily verify that (\ref{ea4}) is correct. It
must be stressed that (\ref{ea4}) is not a fortuitous empirical
coincidence, but rather is a result of using (\ref{ea3}), which
again as noted, can be deduced from
theory.\\
(\ref{ea4}) can be rewritten as
$$l = \sqrt{n}l_P$$
\begin{equation}
t = \sqrt{n}t_P\label{ea5}
\end{equation}
wherein we have used (\ref{ea1}) and (\ref{ea3}) and $n = \sqrt{N}$.\\
We will now compare (\ref{ea5}) with the well known relations, referred to earlier,
\begin{equation}
R = \sqrt{N} l \quad T = \sqrt{N} t\label{ea6}
\end{equation}
The first relation of (\ref{ea6}) is the well known Weyl-Eddington
formula referred to while the second relation of (\ref{ea6}) is
given also on the right side of (\ref{ea3}). We now observe that
(\ref{ea6}) can be seen to be the result of a Brownian Walk
process, $l,t$ being typical intervals between "steps"
(Cf.\cite{uof,rr12,rr13}). We demonstrate this below after equation
(\ref{ea8}). On the other hand, the typical intervals $l,t$ can be
seen to result from a diffusion  process themselves. Let us
consider the well known diffusion relation,
\begin{equation}
(\Delta x)^2 \equiv l^2 = \frac{\hbar}{m} t \equiv \frac{\hbar}{m}
\Delta t\label{ea7}
\end{equation}
(Cf.\cite{rr12},\cite{rr14}-\cite{rr17}). What is being done here is that we are modeling fuzzy spacetime by a double Wiener process to be touched upon later, which leads to (\ref{ea7}). This will be seen in more detail, below.\\
Indeed as $l$ is the Compton wavelength, (\ref{ea7}) can be
rewritten as the Quantum Mechanical Uncertainty Principle
$$l \cdot p \sim \hbar$$
at the Compton scale (Cf. also \cite{rr18}) (or even at the de
Broglie
scale).\\
What (\ref{ea7}) shows is that a Brownian process defines
the Compton scale while (\ref{ea6}) shows that a Random Walk
process with the Compton scale as the interval defines the length
and time scales of the universe itself (Cf.\cite{rr13}). Returning
now to (\ref{ea5}), on using (\ref{ea2}), we observe that in
complete analogy with (\ref{ea7}) we have the relation
\begin{equation}
(\Delta x)^2 \equiv l^2_P = \frac{\hbar}{m_P} t_P \equiv
\frac{\hbar}{m_P} \Delta t\label{ea8}
\end{equation}
We can now argue that the Brownian process (\ref{ea8})
defines the Planck length while a Brownian Random Walk process
with the Planck scale as the interval leads to (\ref{ea5}), that is
the
Compton scale.\\
To see all this in greater detail, it may be observed that
equation (\ref{ea8}) (without subscripts)
\begin{equation}
(\Delta x)^2 = \frac{\hbar}{m} \Delta t\label{eaa}
\end{equation}
is the same as the equation (\ref{ea7}), indicative of a double
Wiener process. Indeed as noted by several scholars, this defines
the fractal Quantum path of
dimension 2 (rather than dimension 1) (Cf.e.g. ref.\cite{rr15}).\\
Firstly it must be pointed out that equation (\ref{eaa}) defines a
minimum space time unit - the Compton scale $(l,t)$. This follows
from (\ref{eaa}) if we substitute into it $\langle \frac{\Delta x}
{\Delta t}\rangle_{max} = c$. If the mass of the particle is the
Planck mass, then this Compton scale becomes the Planck scale.\\
Let us now consider the distance traversed by a particle with the
speed of light through the time interval $T$. The distance $R$
covered would be
\begin{equation}
\int dx = R = c \int dt = cT\label{eIa}
\end{equation}
by conventional reasoning. In view of the equation
(\ref{eaa}), however we would have to consider firstly, the minimum
time interval $t$ (Compton or Planck time), so that we have
\begin{equation}
\int dt \to nt\label{eIIa}
\end{equation}
Secondly, because the square of the space interval $\Delta x$
(rather than the interval $\Delta x$ itself as in conventional
theory) appears in (\ref{eaa}), the left side of (\ref{eIa})
becomes, on using (\ref{eIIa})
\begin{equation}
\int dx^2 \to \int (\sqrt{n}dx) (\sqrt{n}dy)\label{eIIIa}
\end{equation}
Whence for the linear dimension $R$ we would have
\begin{equation}
\sqrt{n}R = nct \quad \mbox{or} \quad R = \sqrt{n} l\label{eIVa}
\end{equation}
Equation (\ref{eIIIa}) brings out precisely the fractal dimension
$D = 2$ of the Brownian path while (\ref{eIVa}) is identical to
(\ref{ea4}) or (\ref{ea6}) (depending on whether we are dealing with
minimum intervals of the Planck scale or Compton scale of
elementary particles). Apart from showing the Brownian character
linking equations (\ref{ea4}) and (\ref{eaa}), incidentally, this
also provides the justification for what has so far been
considered to be a mysterious large number coincidence viz. the
Eddington
formula (\ref{ea6}).\\
There is another way of looking at this. It is well known that in approaches like that of the author or
Quantum Super String Theory, at the Planck scale we have a non
commutative geometry encountered earlier \cite{rr19,rr22}
Indeed as noted, a noncommutative geometry follows without recourse to Quantum
Super Strings, merely by the fact that $l_P,t_P$ are the absolute
minimum space
time intervals as we saw earlier.\\
The non commutative geometry, as is known, is
symptomatic of a modified uncertainty principle at this scale
\cite{rr22}-\cite{rr28}
\begin{equation}
\Delta x \approx \frac{\hbar}{\Delta p} + l^2_P \frac{\Delta
p}{\hbar}\label{ea10}
\end{equation}
The relation (\ref{ea10}) would be true even in Quantum Gravity.
The extra or second term on the right side of (\ref{ea10}) expresses the well known duality effect - as we attempt to go down
to the Planck scale, infact we are lead to the larger scale. The question is, what is this larger scale? If
we now use the fact that $\sqrt{n}$ is the fluctuation in the
number of Planck particles (exactly as $\sqrt{N}$ was the
fluctuation in the particle number as in (\ref{ea3})) so that
$\sqrt{n}mpc = \Delta p$ is the fluctuation or uncertainty in the
momentum for the second term on the right side of (\ref{ea10}), we
obtain for the uncertainty in length,
\begin{equation}
\Delta x = l^2_P \frac{\sqrt{n}m_Pc}{\hbar} =
l_P\sqrt{n},\label{ea11}
\end{equation}
We can easily see that (\ref{ea11}) is the same as the first
relation of (\ref{ea5}). The second relation of (\ref{ea5}) follows
from an
application of the time analogue of (\ref{ea10}).\\
Thus the impossibility of going down to the Planck scale because
of (\ref{ea10}), manifests itself in the fact that as
we attempt to go down to the Planck scale, we infact end up at the
Compton scale. In the next section we will give another demonstration of this result. This is how the Compton scale is encountered in real life.\\
Interestingly while at the Planck length, we have a life time of
the order of the Planck time, as noted above it is possible to
argue on the other hand that with the  mass and length of a
typical elementary particle like the pion, at the Compton scale,
we have a life time which is the age of the universe itself as
shown by
Sivaram \cite{rr3,rr29}.\\
Interestingly also Ng and Van Dam deduce the relations like
\cite{rr30}
\begin{equation}
\delta L \leq (Ll^2_P)^{1/3}, \delta T \leq
(Tt^2_P)^{1/3}\label{ea9}
\end{equation}
where the left side of (\ref{ea9}) represents the uncertainty in the measurement
of length and time for an interval $L,T$. We would like to point
out that if in the above we use for $L,T$, the size and age of
the universe, then $\Delta L$ and
$\Delta T$ reduce to the Compton scale $l,t$.\\
In conclusion, Brownian double Wiener processes and the modification
of the Uncertainity Principle at the Planck scale lead to the
physical Compton scale.
\section{The Universe as Planck Oscillators}
In the previous section, we had argued that a typical
elementary particle like a pion could be considered to be the result of
$n \sim 10^{40}$ evanescent Planck scale particles.  The argument was based on
random motions and also on the modification to the Uncertainity Principle.
We will now consider the problem from a different point of view,
which not only reconfirms the above result, but also enables an elegant
extension to the case of the entire universe itself.
Let us consider an array of $N$ particles, spaced a distance $\Delta x$
apart, which behave like oscillators, that is as if they were connected by
springs. We then have \cite{r2,r3}
\begin{equation}
r  = \sqrt{N \Delta x^2}\label{e1d}
\end{equation}
\begin{equation}
ka^2 \equiv k \Delta x^2 = \frac{1}{2}  k_B T\label{e2d}
\end{equation}
where $k_B$ is the Boltzmann constant, $T$ the temperature, $r$ the extent  and $k$ is the
spring constant given by
\begin{equation}
\omega_0^2 = \frac{k}{m}\label{e3d}
\end{equation}
\begin{equation}
\omega = \left(\frac{k}{m}a^2\right)^{\frac{1}{2}} \frac{1}{r} = \omega_0
\frac{a}{r}\label{e4d}
\end{equation}
We now identify the particles with Planck masses, set $\Delta x \equiv a =
l_P$, the Planck length. It may be immediately observed that use of
(\ref{e3d}) and (\ref{e2d}) gives $k_B T \sim m_P c^2$, which ofcourse agrees
with the temperature of a black hole of Planck mass. Indeed, as noted, Rosen had shown that a Planck mass particle at the Planck scale  can be considered to be a
universe in itself. We also use the fact alluded to that  a typical elementary particle
like the pion can be considered to be the result of $n \sim 10^{40}$ Planck
masses. Using this in (\ref{e1d}), we get $r \sim l$, the pion
Compton wavelength as required. Further, in this latter case, using (48) and the fact that $N = n \sim 10^{40}$, and (\ref{e2d}),i.e. $k_BT = kl^2/N$ and  (\ref{e3d}) and
(\ref{e4d}), we get for a pion, remembering that $m^2_P/n = m^2,$
$$k_ B T = \frac{m^3 c^4 l^2}{\hbar^2} = mc^2,$$
which of course is the well known formula for the Hagedorn temperature for
elementary particles like pions. In other words, this confirms the conclusions
in the previous section, that we can treat an elementary particle as a series of some
$10^{40}$ Planck mass oscillators. However it must be observed from
(\ref{e2d}) and (\ref{e3d}), that while the Planck mass gives the highest
energy state, an elementary particle like the pion is in the lowest energy
state. This explains why we encounter elementary particles, rather than
Planck mass particles in nature. Infact as already noted, a Planck
mass particle decays via the Bekenstein radiation within a Planck time
$\sim 10^{-42}secs$. On the other hand, the lifetime of an elementary particle
would be very much higher.\\
In any case the efficacy of our above oscillator model can be seen by the fact that we recover correctly the masses and Compton scales in the order of magnitude sense and also get the correct Bekenstein and Hagedorn formulas as seen above, and get the correct estimate of the mass of the universe itself, as will be seen below.\\
Using the fact that the universe consists of $N \sim 10^{80}$ elementary
particles like the pions, the question is, can we think of the universe as
a collection of $n N \, \mbox{or}\, 10^{120}$ Planck mass oscillators? This is what we will now
show. Infact if we use equation (\ref{e1d}) with
$$\bar N \sim 10^{120},$$
we can see that the extent $r \sim 10^{28}cms$ which is of the order of the diameter of the
universe itself. Next using (\ref{e4d}) we get
\begin{equation}
\hbar \omega_0^{(min)} \langle \frac{l_P}{10^{28}} \rangle^{-1} \approx m_P c^2 \times 10^{60} \approx Mc^2\label{e5d}
\end{equation}
which gives the correct mass $M$, of the universe which in contrast to the earlier pion case, is the highest energy state while the Planck oscillators individually are this time the lowest in this description. In other words the universe
itself can be considered to be described in terms of normal modes of Planck scale oscillators.\\
We will return to these considerations later: this and the preceeding considerations merely set the stage.
\section{Fuzzy Spacetime}
Thus we are lead to a minimum scale which is the Compton scale in the real world. Once such a minimum spacetime scale is invoked, then we have a non commutative geometry as shown by Snyder more than fifty years ago \cite{snyder}:
$$[x,y] = (\imath a^2/\hbar )L_z, [t,x] = (\imath a^2/\hbar c)M_x, etc.$$
\begin{equation}
[x,p_x] = \imath \hbar [1 + (a/\hbar )^2 p^2_x];\label{e3z}
\end{equation}
The relations (\ref{e3z}) are compatible with Special Relativity. Indeed such minimum spacetime models were studied for several decades, precisely to overcome the divergences encountered in Quantum Field Theory \cite{x1}-\cite{x9}.\\
Before proceeding further, it may be remarked that when the square of a, which we will take to be the Compton wavelength (including the Planck scale, which is a special case of the Compton scale for a Planck mass viz., $10^{-5}gm$), in view of the above comments  can be neglected, then we return to point Quantum Theory.\\
It is interesting that starting from the usual Dirac coordinate we can deduce the non commutative geometry (\ref{e3z}), independently. For this we note that the $\alpha$'s in the Dirac coordinate are given by
$$\vec \alpha = \left[\begin{array}{ll}
\vec \sigma \quad 0\\
0 \quad \vec \sigma
\end{array}
\right]\quad \quad ,$$
the $\sigma$'s being the Pauli matrices.
We next observe that the first term on the right hand side is the usual Hermitian position. For the second term which contains $\alpha$, we can easily verify from the commutation relations of the $\sigma$'s that
\begin{equation}
[x_\imath , x_j] = \beta_{\imath j} \cdot l^2\label{eA}
\end{equation}
where $l$ is the Compton scale.\\
There is another way of looking at this. Let us consider the Dirac coordinate to be complex, reminiscent of Newman's original complexification of the coordinate. We now try to generalize this complex coordinate to three dimensions. Then we encounter a surprise - we end up with not three, but four dimensions,
$$(1, \imath) \to (I, \sigma),$$
where $I$ is the unit $2 \times 2$ matrix. We get the special relativisitc Lorentz invariant metric at the same time. (In this sense, as noted by Sachs \cite{sachs}, Hamilton who made this generalization would have also hit upon special relativity, if he had identified the fourth coordinate with time).\\
That is,\\
$$x + \imath y \to Ix_1 + \imath x_2 + jx_3 + kx_4,$$
where $(\imath ,j,k)$ now represent the Pauli matrices; and, further,
$$x^2_1 + x^2_2 + x^2_3 - x^2_4$$
is invariant.\\
While the usual Minkowski four vector transforms as the basis of the four dimensional representation of the Poincare group, the two dimensional representation of the same group, given by the right hand side in terms of Pauli matrices, obeys the quaternionic algebra of the second rank spinors (Cf.Ref.\cite{sakharov,heine,bgsfpl} for details).\\
To put it briefly, the quarternion number field obeys the group property and this leads to a number system of quadruplets as a minimum extension. In fact one representation of the two dimensional form of the quarternion basis elements is the set of Pauli matrices. Thus a quarternion may be expressed in the form
$$Q = -\imath \sigma_\mu x^\mu = \sigma_0x^4 - \imath \sigma_1 x^1 - \imath \sigma_2 x^2 - \imath \sigma_3 x^3 = (\sigma_0 x^4 + \imath \vec \sigma \cdot \vec r)$$
This can also be written as
$$Q = -\imath \left(\begin{array}{ll}
\imath x^4 + x^3 \quad x^1-\imath x^2\\
x^1 + \imath x^2 \quad \imath x^4 - x^3
\end{array}\right).$$
As can be seen from the above, there is a one to one correspondence between a Minkowski four-vector and $Q$. The invariant is now given by $Q\bar Q$, where $\bar Q$ is the complex conjugate of $Q$.\\
However, as is well known, there is a lack of spacetime reflection symmetry in this latter formulation. If we require reflection symmetry also, we have to consider the four dimensional representation,
$$(I, \vec \sigma) \to \left[\left(\begin{array}{ll}
I \quad 0 \\
0 \quad -I
\end{array}\right), \left(\begin{array}{ll}
0 \quad \vec \sigma \\
\vec \sigma \quad 0
\end{array}\right)\right] \equiv  (\Gamma^\mu)$$
(Cf.also.ref. \cite{heine} for a detailed discussion). The motivation for such a reflection symmetry is that usual laws of physics, like electromagnetism do indeed show the symmetry.\\
 We at once deduce spin and special relativity and the geometry (\ref{e3}). This is a transition that has been long overlooked \cite{sakharov}. Conversely it must be mentioned that spin half itself is relational and refers to three dimensions, to a spin network infact \cite{spin}. That is, spin half is not meaningful in a single particle universe.\\
Equally interesting is the fact that starting from the geometry (\ref{e3}) we can deduce the Dirac equation itself.\\
While a relation like (\ref{eA}) above has been in use recently, in non commutative models, and as noted, was an independent starting point due to the work of Snyder, we would like to stress that it has been overlooked that the origin of this non commutativity lies in the original Dirac coordinates.\\
The above relation shows on comparison with the position-momentum commutator that the coordinate $\vec x$ also behaves like a ``momentum''. This can be seen directly from the Dirac theory itself where we have \cite{uof}
\begin{equation}
c\vec \alpha = \frac{c^2\vec p}{H} - \frac{2\imath}{\hbar} \hat x H\label{ea}
\end{equation}
In (\ref{ea}), the first term is the usual momentum. The second term is the extra ``momentum'' $\vec p$ due to zitterbewegung.\\
Infact we can easily verify from (\ref{ea}) that
\begin{equation}
\vec p = \frac{H^2}{\hbar c^2}\hat x\label{eb}
\end{equation}
where $\hat x$ has been defined in (\ref{ea}).\\
We finally investigate what the angular momentum $\sim \vec x \times \vec p$ gives - that is, the angular momentum at the Compton scale. We can easily show that
\begin{equation}
(\vec x \times \vec p)_z = \frac{c}{E} (\vec \alpha \times \vec p)_z = \frac{c}{E} (p_2\alpha_1 - p_1\alpha_2)\label{ec}
\end{equation}
where $E$ is the eigen value of the Hamiltonian operator $H$. Equation (\ref{ec}) shows that the usual angular momentum but in the context of the Compton scale, leads to the ``mysterious'' Quantum Mechanical spin.\\
In the above considerations, we started with the Dirac equation and deduced the underlying non commutative geometry of space time. Interestingly, starting with Snyder's non commutative geometry, based solely on Lorentz invariance and a minimum spacetime length, which we have taken to be the Compton scale, (\ref{e3}),
it is possible to deduce the relations (\ref{ec}), (\ref{eb}) and the Dirac equation itself as noted \cite{bgskluwer2} and as we will see shortly.\\
We have thus established the correspondence between considerations starting from the Dirac theory of the electron and Snyder's (and subsequent) approaches based on a minimum spacetime interval and Lorentz covariance. It can be argued from an alternative point of view that special relativity operates outside the Compton wavelength (Cf.ref.\cite{uof}).\\
We could have started with the Kerr-Newman Black Hole. Infact the derivation of the Kerr-Newman Black Hole itself begins with a Quantum Mechanical spin yielding complex shift, which Newman has found inexplicable even after several decades \cite{newman,et}. As he observed, ``...one does not understand why it works. After many years of study I have come to the conclusion that it works simply by accident''. And again, ``Notice that the magnetic moment $\mu = ea$ can be thought of as the imaginary part of the charge times the displacement of the charge into the complex region... We can think of the source as having a complex center of charge and that the magnetic moment is the moment of charge about the center of charge... In other words the total complex angular momentum vanishes around any point $z^a$ on the complex world-line. From this complex point of view the spin angular momentum is identical to orbital, arising from an imaginary shift of origin rather than a real one... If one again considers the particle to be ``localized'' in the sense that the complex center of charge coincides with the complex center of mass, one again obtains the Dirac gyromagnetic ratio...''\\
The unanswered question has been, why does a complex shift somehow represent spin about that axis? The answer to this question lies in the above considerations. Complexified space time is symptomatic of fuzzy space time and a non commutative geometry and Quantum Mechanical spin \cite{fpl}. Indeed Zakrzewski has shown in a classical context that non commutativity implies spin \cite{zak,bgschubykalo}.\\
The above considerations used the Quantum Mechanical spin together with classical relativity, though the price to pay for this was minimum space time intervals and noncommutative geometry. Is this the path towards a reconciliation of electromagnetism and gravitation?
\section{Discussion}
In String Theory we have the vibration of one dimensional strings at the Planck scale with mechanical concepts like tension, playing a role. In the case of Planck oscillators, we have a one dimensional oscillator associated with the background Zero Point Field or Dark Energy. In String Theory the different vibrational modes corresponds to different particles. In the latter case collective modes of one dimensional oscillators constitute an elementary particle at the Compton scale. Moreover this leads to noncommutative geometry and fuzzy spacetime at the Compton scale $l, \tau$ with a quantum of area. If $O(l^2)$ is neglected we get back ordinary Quantum Mechanics, or retaining $\sim l^2$ we get a rationale for the Dirac equation. Moreover if fluctuations are considered out of the background Zero Point Field we recover a cosmology with acceleration and a small cosmological constant, which goes over to the Big Bang cosmology in the limit $l \to 0$. In String Theory on the other hand, there is no such limit to conventional physics and moreover the cosmological constant is some $10^{120}$ times its observed value. Moreover there are extra dimensions - ten or eleven as in the latest version of M-Theory. No such extra dimensions are required in the latter theory. String Theory finds a place for the graviton while the latter theory recovers linearized gravitation from the noncommutative structure and gives a distributional picture of gravity, one in which gravitation is not fundamental \cite{uof}.\\
So both theories originate in one dimension at the Planck Scale, but the latter theory goes on to describe the universe at the Planck scale as a large polymer, whereas the description is different at the Compton scale. In fact at this scale we encounter two dimensions and a breakdown of the coherence seen in the Planck scale polymer. At even greater distances we encounter the three dimensional physical space \cite{gip}. Moreover the picture is Machian and all the hitherto supposedly mysterious coincidences are actually consequences of the theory. Furthermore there is a mass spectrum for all the known elementary particles and other experimental effects \cite{uof}.


\begin{thebibliography}{99}
\bibitem {reg} Regge, T., and Alfaro, V. de., ``Potential Scattering'', North Holland Publishing Co., Amsterdam, 1965.
\bibitem {roma} Roman, P., ``Advanced Quantum Theory'', Addison-Wesley, Reading, Mass., 1965, p.31.
\bibitem {tassie} Bogdan, P., Rith, K., ``Particles and Nuclei: An Introduction to the Physical Concepts'', Springer-Verlag, Berlin, 1993.
\bibitem {rohr} Rohrlich, F., ``Classical Charged Particles'', Addison-Wesley, Reading, Mass., 1965.
\bibitem {barut} Barut, A.O., ``Electrodynamics and Classical Theory of Fields and Particles'', Dover Publications, Inc., New York, 1964, p.97ff.
\bibitem {feynman} Feynman, R.P., Leighton, R.B., and Sands, M., ``The Feynman Lectures on Physics'', Vol.II, Addison-Wesley Publishing Co., Inc., Mass., 1965.
\bibitem {dirac} Dirac, P.A.M., Proc.Roy.Soc., London, A268, 1962, p.57.
\bibitem {ven} Veneziano, G., in ``The Geometric Universe'', Ed. by S.A. Huggett, et al., Oxford University Press, Oxford, 1998, p.235ff.
\bibitem {venezia} Veneziano, G., Physics Reports, 9, No.4, 1974, p.199-242.
\bibitem {lee} Lee, T.D., ``Particle Physics and Introduction to Field Theory'', Harwood Aademic, 1981, pp.391ff.
\bibitem {smolin} Martin, B.R., Shaw, G., ``Particle Physics'', John Wiley \& Sons, New York, 1992.
\bibitem {walter} Sidharth, B.G., ``Fuzzy, non commutative spacetime: A new paradigm for a new century'' in Proceedings of Fourth International Symposium on ``Frontiers of Fundamental Physics'', Kluwer Academic/Plenum Publishers, New York, 2001, p.97-108.
\bibitem {fog} Fogleman, G., Am.J.Phys., 55 (4), 1987, p.330-6.
\bibitem {bgskluwer} Schwarz, J., Physics Reports, 89 (227), 1982.
\bibitem {st} Jacob, M., Ed., Physics Reports, Reprint Volume, North-Holland, Amsterdam, 1974.
\bibitem {veneziano} Veneziano, G., Quantum geometric origin of all forces in string theory. In: Huggett SA, et al., editors. ``The Geometric Universe'', Oxford University Press, Oxford, 1988, p.235-43.
\bibitem {ramond} Ramond, P., Phys.Rev.D., 3 (10), 1971, pp.2415-2418.
\bibitem {kaluza} Kaluza, Th., in ``An Introduction to Kaluza-Klein Theories'', World Scientific, Singapore, 1984.
\bibitem {witten} Witten, E., Physics Today, April 1996, pp.24-30.
\bibitem {olive} Olive, D.I., Nulear Phys.B (Proc. Suppl), 46, 1996, p.1-15.
\bibitem {green} Greene, B.,  ``The Elegant Universe'', Vintage, London, 1999, p.15.
\bibitem {madore} Madore, J., ``An Introduction to Non-Commutative Differential Geometry'', Cambridge University Press, Cambridge, 1995.
\bibitem {bgsafdb} Sidharth, B.G.,  Annales Fondation L. De Broglie, 29 (3), 2004, 1.
\bibitem {rr1} Kempf, A., in ``From the Planck length to the Hubble radius'', Ed. A. Zichichi, World Scientific, Singapore, 2000, pp.613ff.
\bibitem {rr2} Veneziano, G.,  in ``The Geometric Universe'', Ed. by S.A.
Huggett, et al., Oxford University Press, Oxford, 1998, p.235ff.
\bibitem {garay} Garay, L.J.,  Int.J.Mod.Phys.A., 1995, 10 (2), 145-165.
\bibitem {rr3} Sivaram, C., Am.J.Phys., 51(3), 1983, p.277.
\bibitem {rr5} Rosen, N., Int.J.Th.Phys., 32(8), 1993, p.1435-1440.
\bibitem {rr6} Sidharth, B.G.,  and Popova, A.D., Differential Equations
and Dynamical Systems, 4 (3/4), 1996, 431-440.
\bibitem {uof} Sidharth, B.G., ``Universe of Fluctuations'', Springer, Netherlands, 2005.
\bibitem {rr8} Sidharth, B.G., Int.J.Mod.Phys.A, 13 (15), 1998, p.2599ff.
\bibitem {rr9} Sidharth, B.G., Int.J.Th.Phys., 37 (4), 1998, p.1307ff.
\bibitem {rr10} Sidharth, B.G., Il Nuovo Cimento, 115B (12), (2), 2000, pg.151.
\bibitem {rr11} Norman, E.G., Am.J.Phys., 544, 317, 1986.
\bibitem {melnikov} Melnikov, V.N.,  Int.J.Th.Phys., \underline{33} (7), 1994, 1569-1579.
\bibitem {rr12} Sidharth, B.G., Chaos, Solitans and Fractals, 12(1), 2001,
173-178.
\bibitem {rr13} Sidharth, B.G., Chaos,Solitons and Fractals, 13, 2002, pp.1325-1330.
\bibitem {rr14} Nelson, E., Physical Review, Vol.150, No.4, October 1966, p.1079-1085.
\bibitem {rr15} Nottale, L., ``Fractal Space-Time and Microphysics: Towards
a Theory of Scale Relativity'',World Scientific, Singapore,
1993, p.312.
\bibitem {rr16} Smolin, L.,  in ``Quantum Concepts in Space and Time'', Eds.,
R. Penrose and C.J. Isham, OUP, Oxford, 1986, pp.147-181.
\bibitem {rr17} Kyprianidis, A., Int.J.Th.Phys., 1992, pp.1449-1483.
\bibitem {rr18} Sidharth, B.G., ``Concise Encyclopaedia of SuperSymmetry
and Non Commutative Structures in Mathematics and Physics'', Eds. J.
Bagger, S. Duply, W. Sugel, New York, Kluwer Academic Publishers,
2001.
\bibitem {rr19} Ne'eman, Y., in Proceedings of the First International Symposium,
``Frontiers of Fundamental Physics'', Eds. B.G. Sidharth and A.
Burinskii, Universities Press, Hyderabad, 1999, pp.83ff.
\bibitem {rr22} Amati, D., in ``Sakharov Memorial Lectures'', Eds. L.V. Kaddysh
and N.Y. Feinberg, Nova Science, New York, 1992, pp.455ff.
\bibitem {rr23} Guida, R., et al.,  Mod.Phys.Lett. A, Vol.6, No.16, 1991, p.1487-1503.
\bibitem {rr24} Witten, W., Physics Today, April 1996, pp.24-30.
\bibitem {rr25} Garay, L.J., Int.J.Mod.Phys. A, Vol.10, No.2, 1995, p.145-165.
\bibitem {rr26} Maggiore, M.,  Phys.Lett.B., 319, 1993, p.83-86.
\bibitem {rr27} Maggiore, M.,  Phys.Rev.D., Vol.49, No.10, 1994, p.5182-5187.
\bibitem {rr28} Doplicher, S., et al., Phys.Lett., B, 331, 1994, p.39-44.
\bibitem {rr29} Sivaram, C.,  Am.J.Phys., 50(3), 1982, p.279ff.
\bibitem {rr30} Van Dam, H., et al.,  Mod.Phys.Lett.,A, Vol.9, No.4, 1994, 335-340.
\bibitem {r2} Sidharth, B.G., Found.Phys.Lett., 15 (6), 2002, pp.577-583.
\bibitem {r3} Goodstein, D.L.,  ``States of Matter'', Dover Publications Inc., New York, 1985, p.160ff.
\bibitem {snyder} Snyder, H.S., Physical Review, Vol.72, No.1, July 1 1947, p.68-71.
\bibitem {x1} Snyder, H.S.,  Physical Review, Vol.72, No.1, January 1 1947, p.38-41.
\bibitem {x2} Kardyshevskii, V.G.,  Translated from Doklady Akademii Nauk SSSR, Vol.147, No.6, December 1962, p.1336-1339.
\bibitem {x3} Wolf, C.,  Hadronic Journal, Vol.13, 1990, p.22-29.
\bibitem {x4} Hooft, G. 't.,  ArXiv:gr-qc/9601014.
\bibitem {x5} Bombelli, L.,  Lee, J., Meyer, D., and  Sorkin, R.D.,  Physical Review Letters, Vol.59, No.5, August 1987, p.521-524.
\bibitem {x6} Crane, L.,  and Smolin, L.,  Nucl.Phys., B267, 1986, p.714.
\bibitem {x7} Caldirola, P.,  Supplemento Al Volume III, Serie X, del Nuovo Cimento, No.2, 1956, p.297ff.
\bibitem {x8} Schild, A.,  Phys.Rev., 73, 1948, p.414-415.
\bibitem {x9} Lee, T.D.,  Physics Letters, Vol.122B, No.3,4, 10 March 1983, p.217-220.
\bibitem {sachs} Sachs, M., ``General Relativity and Matter'', D. Reidel Publishing Company, Holland, 1982, p.45ff.
\bibitem {sakharov} Sidharth, B.G., Found.Phys.Lett., 16 (1), 2003, pp.91-97.
\bibitem {heine} Heine, V.,  ``Group Theory in Quantum Mechanics'', Pergamon Press, Oxford, 1960, p.364.
\bibitem {bgsfpl} Shirokov, Yu. M., Soviet Physics JETP \underline{6} (33), No.5, 1958, pp.929-935.
\bibitem {spin} Penrose, R.,  ``Angular Momentum: An approach to combinational space-time'' in, ``Quantum Theory and Beyond'', Ed., Bastin, T., Cambridge University Press, Cambridge, 1971, pp.151ff.
\bibitem {bgskluwer2} Sidharth, B.G.,  in Proceedings of Fourth International Symposium on ``Frontiers of Fundamental Physics'', Kluwer Academic, New York, 2001, pp.97-108.
\bibitem {newman} Newman, E.T.,  J.Math.Phys., 14 (1), 1973, p.102.
\bibitem {et} Newman, E.T.,  Enrico Fermi International School of Physics
1975, Proceedings p557.
\bibitem {fpl} Sidharth, B.G.,  Foundation of Physics Letters, {\bf 16} (1), 91-97 (February 2003).
\bibitem {zak} Zakrewski, S., ``Quantization, Cohrent States and Complex Structures'', Ed. by J.P. Antoine et al, Plenum Press, New York, 1995, p.249ff.
\bibitem {bgschubykalo} Sidharth, B.G.,  ``An Interface between Classical Electrodynamics and Quantum Mechanics'', in ``Has the last word been said on Classical Electrodynamics?'', Eds. A. Chubykalo et al., Rinton Press, USA, 2003.
\bibitem {gip} Sidharth, B.G., ``The Counter intuitive Universe'' in, ``A Century of Ideas'', Springer, Netherlands, 2006 (in press).
\end{thebibliography}
\end{document}